# The role of double $TiO_2$ layers at the interface of FeSe/ $SrTiO_3$ superconductors

Ke Zou[1, 2], Subhasish Mandal[1, 2], Stephen Albright[2, 3], Rui Peng[4], Yujia Pu[4], Divine Kumah[1, 2], Claudia Lau[2, 3], Georg Simon[2, 5], Omur E. Dagdeviren[2, 5], Xi He[1,6], Ivan Božović[1,6], Udo D. Schwarz[2, 5, 7], Eric I. Altman[2, 7], Donglai Feng[4], Fred J. Walker[1, 2], Sohrab Ismail-Beigi[1, 2], and Charles H. Ahn[1, 2, 3, 5]

[1] *Department of Applied Physics, Yale University, New Haven CT 06520, USA*

[2] *Center for Research on Interface Structures and Phenomena (CRISP), Yale University, New Haven CT 06520, USA*

[3] *Department of Physics, Yale University, New Haven CT 06520, USA*

[4] *State Key Laboratory of Surface Physics, Department of Physics and Advanced Materials Laboratory, Fudan University, Shanghai 200433, China*

[5] *Department of Mechanical Engineering and Materials Science, Yale University, New Haven CT 06520, USA*

[6] *Condensed Matter Physics and Materials Science Department, Brookhaven National Laboratory, Upton, NY 11973, USA*

[7]*Department of Chemical and Environmental Engineering, Yale University, New Haven CT 06520, USA*

**We determine the surface reconstruction of $SrTiO_3$ used to achieve superconducting FeSe films in experiments, which is different from the 1x1 $TiO_2$ terminated $SrTiO_3$ assumed by most previous theoretical studies. In particular, we identify the existence of a double $TiO_2$ layer at the $SrTiO_3$-FeSe interface that plays two important roles. First, it facilitates the epitaxial growth of FeSe. Second, *ab initio* calculations reveal a strong tendency for electrons to transfer from an oxygen deficient $SrTiO_3$ surface to FeSe when the double $TiO_2$ layer is present. As a better electron donor than previously proposed interfacial structures, the double layer helps to remove the hole pocket in the FeSe at the Γ point of the Brillouin zone and leads to a band structure characteristic of superconducting samples. The characterization of the interface structure presented here is a key step towards the resolution of many open questions about this novel superconductor.**

The discovery of the iron based superconductors (SC) [1-14] exhibiting unconventional superconductivity promises to enhance our understanding of superconductivity and lead to the development of new materials with high critical temperature ($T_c$). Iron pnictides [15-18] and chalcogenides [1-14, 19-34], the simplest being LiFeAs ($T_c$ = 18 K) [35-37] and FeSe ($T_c$ = 9 K) [1], respectively, share important similarities with cuprate superconductors [38, 39]. Both possess a two-dimensional layered structure and exhibit antiferromagnetic ordering or other ordering (such as nematic ordering) at low temperatures that competes with superconducting phases. They also show enhanced superconductivity when dopants are added to their layered structures (e.g., hole doping of CuO planes in $YBa_2Cu_3O_{7-x}$ [39], doping of FeAs planes in $SmFeAsO_{1-x}F_x$ [17], K doping in bulk planar FeSe superconductors [32]).

A particularly intriguing Fe-based SC is one-unit-cell (1UC) FeSe grown on $SrTiO_3$ (STO) substrates via molecular beam epitaxy, with a reported [12] $T_c$ above 100 K, the highest among Fe-based SCs and rivaled only by a few cuprates [38, 39] and $H_2S$ under extreme pressure [40]. This surprising enhancement of $T_c$ in 1 UC FeSe/STO compared to the bulk phase [41, 42] has been ascribed to interactions between FeSe and STO due to the interfacial nature of the system. 1 UC FeSe grown on BaTiO$_3$ [9] and on anatase TiO$_2$ [43] also shows enhanced Tc. All three substrates are terminated with TiO$_2$ layers. The mechanism for Cooper pairing in this material is an active area of investigation, and a deeper understanding of the interactions between FeSe and STO may provide insight into the origins of unconventional SC.

In this letter, we experimentally determine the atomic structure of 1 UC FeSe/STO and use density functional theory to explore how it affects the electronic structure. Specifically, we identify an STO surface reconstruction that is not a simple truncation of the bulk structure, but represents a change in interface stoichiometry, whereby the STO terminates with two monolayers (2L) of TiO$_2$. We show that this reconstruction is critical in two ways. First, it facilitates the growth of coherently strained, epitaxial FeSe. Second, the 2L TiO$_2$-terminated surface facilitates electron transfer to FeSe by modifying both the electronic structure and defect chemistry at the FeSe-STO interface. The importance of a reconstructed surface has been suggested to be the key to achieving high $T_c$ in this materials system [4, 25, 29, 33], but so far has not been included in theoretical attempts to account for enhanced superconductivity in 1 UC FeSe/STO.

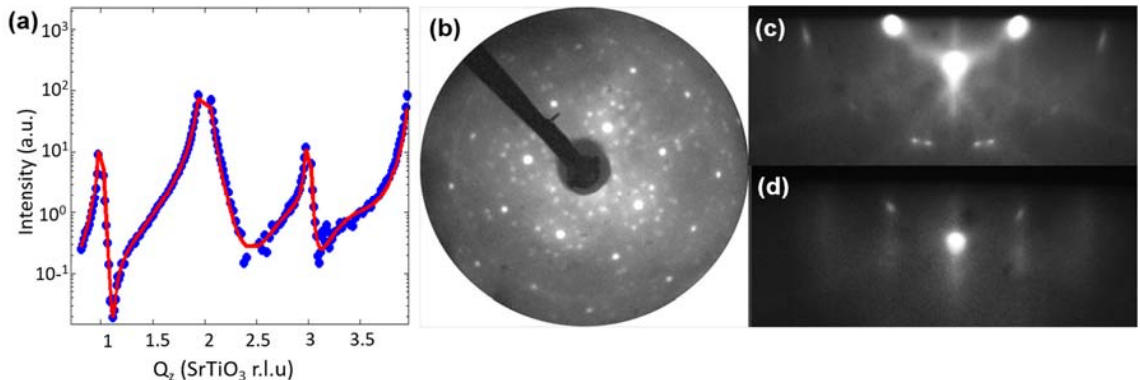

FIG 1. The identification of 2L TiO$_2$ on STO and the epitaxial growth of 1x1 FeSe. (a) (2, 0, $\ell$) crystal truncation rod (CTR) of a treated STO substrate (blue circles). The red line is a fit using a 2L surface model. (b) and (c) Low energy electron diffraction (LEED) and reflection high energy electron diffraction (RHEED) of STO substrates showing the $\sqrt{13} \times \sqrt{13}$ $R33.7°$ reconstruction. (d) RHEED of 1 UC FeSe on STO showing the 1x1 structure of the FeSe.

The growth procedures of the FeSe/STO are similar to those reported in Refs. [5, 8, 10, 12, 25, 29]. The STO substrates are prepared using a high temperature (950 °C) anneal in air at ambient pressure by using the procedures described in Refs. [5, 8]. In addition, bulk STO substrates terminated with a single TiO$_2$ layer are prepared using a standard etch procedure [44]. The substrates are annealed in flowing oxygen at approximately 1000 °C, as described in Section 1 of the Supplementary Information [45].

Similar high temperature annealing usually leads to a 2L-TiO$_2$ surface termination with various reconstructions [46-48]. We confirm the existence of 2L-TiO$_2$ using synchrotron x-ray diffraction. The experiments are performed in vacuum at 300 K at the 33ID beam line at the Advanced Photon Source with an incident x-ray wavelength of 0.83 Å (15 keV). The (2, 0, $\ell$) crystal truncation rod measured for STO is shown in Fig. 1(a), together with a fit to a model of a 2L-TiO$_2$ surface termination. Note that the rods of 1x1 1L-TiO$_2$ terminated STO show features distinct from the 2L-TiO$_2$ terminated STO, as shown in Section 4 of the Supplementary Information [45].

Subsequent experiments reveal that the annealing results in a highly ordered $\sqrt{13} \times \sqrt{13}$ R33.7° reconstruction, easily identified by the sharp diffraction features in synchrotron x-ray diffraction (Fig. 2(a)), low energy electron diffraction (LEED) (Fig. 1(b)), and reflection high energy electron diffraction (RHEED) (Fig. 1(c)). This RHEED pattern is strikingly similar to those shown in prior reports [25, 29, 33] of 1UC FeSe/STO, though the nature of the reconstruction is not identified.

By comparing FeSe growth on STO substrates with the two different types of termination, we establish that a 2L-TiO$_2$ terminated STO surface promotes the epitaxial growth of FeSe. After preparing the 2L-TiO$_2$ surface, samples are loaded for chalcogenide molecular beam epitaxy (MBE) growth. Before growth, the substrates are treated by high temperature vacuum and Se annealing that leads to a less ordered surface reconstruction while retaining the 2L-TiO$_2$ termination. The substrates are cooled to a growth temperature of ~450-500 C°. 1UC FeSe films grown on such reconstructed surfaces are epitaxial and coherently strained, as indicated by RHEED (Fig. 1(d)). Note that the RHEED for the 1 UC FeSe has a 1x1 pattern, which is the same symmetry as the one observed in 1L-TiO$_2$ terminated STO substrates, i.e., it does not show the $\sqrt{13} \times \sqrt{13}$ reconstruction that is apparent in RHEED from the 2L-TiO$_2$ terminated STO substrate just before growth (Fig. 1(c)). However, synchrotron x-ray diffraction indicates that even after deposition of FeSe, the $\sqrt{13} \times \sqrt{13}$ reconstruction of the STO is still present under FeSe (Fig. 2(a)). Growth on surfaces terminated with a single layer of TiO$_2$ under a range of conditions results in films that show diffuse RHEED patterns, indicative of disordered films (Fig. S1 and Fig. S3) [45].

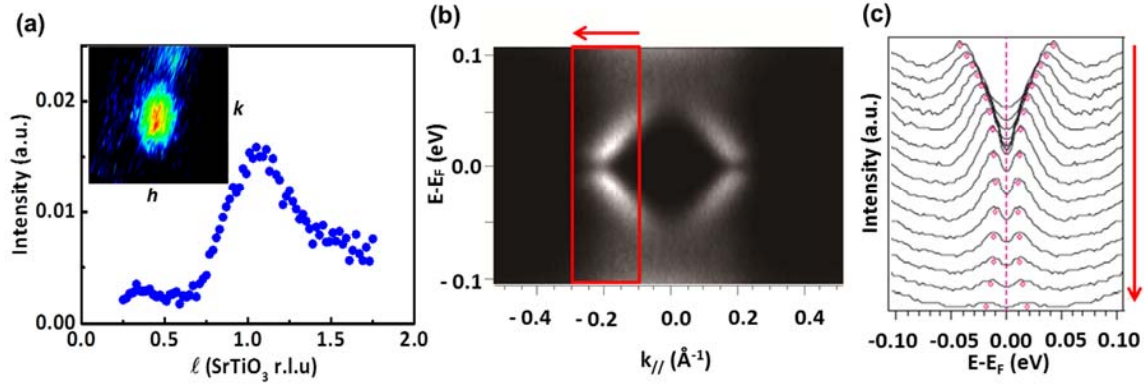

FIG 2. 1UC FeSe/STO with 2L TiO$_2$ at the interface. (a) A crystal truncation rod along (2/13, 10/13, $\ell$) of FeSe/STO. Inset: Reciprocal space mapping around the (2/13, 10/13, 0.8) peak. The ranges of $h$ and $k$ in the inset are 0.1 with a unit of 13 r.l.u.. Both plots indicate that the $\sqrt{13} \times \sqrt{13}$ reconstruction of STO remains beneath FeSe. (b) ARPES intensity near the M point at the corner of the Brillouin zone. To highlight the gap at the Fermi level, the data above the Fermi level are the mirror images of the data below the Fermi level. (c) The symmetrized energy distribution curves along the portion indicated by the red box and arrow in (b).

To examine the electronic structure of the samples, we carry out *in situ*, low temperature ARPES on 1UC FeSe grown on a 2L-TiO$_2$ terminated STO [5, 8]. Data taken near the M point of the Brillouin zone are shown in Fig. 2(b). At T = 25 K, an electron band is observed at the M point, as in other reports [5, 28, 29] of superconducting 1UC FeSe/STO. Figure 2(c) shows symmetrized photoemission spectra (energy distribution curves, EDCs) across the Fermi level for the area marked by the red box in Fig. 2(b). The opening of a gap is indicated by a dip at the Fermi level in the EDCs and by the inflection of the quasiparticle dispersion near the Fermi level, as marked by the red symbols (Fig. 2(c)) [5, 28, 29]. From these data we conclude that a 2L-TiO$_2$ terminated STO substrate supports the electronic structure characteristic of superconducting FeSe.

To explore the role of the 2L-TiO$_2$ interface on the electronic structure of 1UC FeSe/STO, we compute the band structure using density functional theory (DFT) for FeSe on 1L- and 2L-TiO$_2$ terminated STO[45]. A significant feature of the ARPES in superconducting 1UC FeSe, unlike in bulk FeSe, is the absence of a hole pocket at the Γ point and the presence of an electron band at the M point (see Fig. 2(b)) [5, 28, 29]. We find that a 2L-TiO$_2$ terminated STO surface promotes charge transfer and can suppress this feature of the electronic structure more effectively than a 1L-TiO$_2$ termination. The band structures of ideal 1UC FeSe/STO with 1L- and 2L-TiO$_2$ terminations are shown in Figs. 3(a) and 3(b), respectively. Both ideal terminations have similar band structures with a prominent hole pocket predicted around Γ, which is absent in the experiments; electron transfer to the FeSe is required to fill this pocket. The prime candidate source of electrons comes from oxygen vacancies in the STO surface layers [4, 5, 49], a mechanism we explore below, although other defects could play a role as well [50].

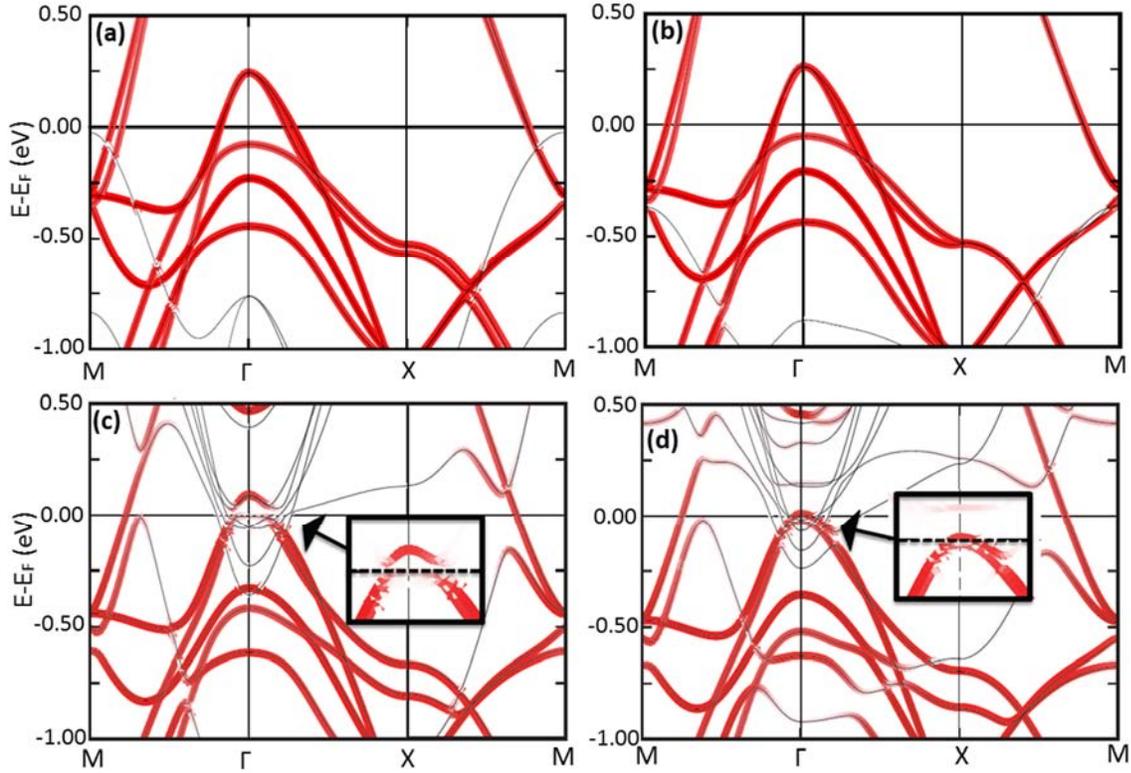

FIG 3. Orbitally resolved band structures for Fe-3$d$ for four models of 1UC FeSe/STO; (a) with stoichiometric 1L-TiO$_2$ termination, (b) with stoichiometric 2L termination, (c) with 1L termination and 50% oxygen vacancies, and (d) with 2L termination and 50% oxygen vacancies. Thin black lines represent the electron energy bands for the whole heterostructure; the red overlay represents projections of the Fe-3$d$ orbital, where the color intensity is proportional to the projection amount. The Fermi level is at zero in all cases. The energy scale is in eV. Insets in (c) and (d): zoom-in on the Fe bands around the Γ-point.

With this motivation, we have performed DFT calculations for interfaces with 50% O vacancies in the surface layer of STO for both the 1L- and 2L-TiO$_2$ terminations. The removal of oxygen results in a significant structural and energetic change at the interface. Similar results are reported in Ref. [4]. The ideal stoichiometric STO surfaces bond weakly to the FeSe, with Ti-Se distances of 3.11 Å and 3.50 Å in the 1L and 2L configurations, along with extremely weak binding energies of 0.073 eV and 0.089 eV per Ti-Se bond. The introduction of O vacancies shortens and greatly strengthens the bonds: the Ti-Se distances become 2.82 Å and 2.67 Å in the 1L and 2L cases, respectively, with sizable binding energies of 0.60 eV and 0.88 eV per Ti-Se bond. The 2L-TiO$_2$ geometry with vacancies displays stronger binding and thus further promotes epitaxy. A similar enhancement of bonding along with smaller Se-Ti distances of 2.6 Å and 2.9 Å in the 1L geometry has been reported [4, 50]. We note all the calculated Se-Ti distances are smaller than the available experimental value [51]. One reason for the discrepancy may be that far fewer O vacancies are present in the experimental system, which will generally reduce charge transfer and decrease the binding.

The stronger binding in the 2L case goes hand in hand with stronger electron doping from STO to FeSe (Figs. S4(a)–(d) show the electron transfer [45]), which moves the FeSe Fermi level up in energy (see Fig. 3(c) and 3(d)). In the 2L termination, this doping removes the hole pocket at the Γ point, recovering the band structure observed in ARPES. For the same vacancy concentration, the 1L system still has a noticeable hole pocket at Γ (compare Figs. 3(c) and 3(d)). One simple reason for the enhanced 2L electron transfer is that the electrons are already more effectively confined near the surface in the case of 2L-$TiO_2$ termination (see Fig. S5 for details [45]), while in the 1L termination, some of the electrons near the surface 'leak back' into the STO substrate and are unavailable for transfer into the FeSe.

We have also computed the consequence of having very high oxygen vacancy concentration of 100% on the 1x1 1L-$TiO_2$ and 2L-$TiO_2$ surfaces. In both cases, the added electrons stay in the STO and do not transfer to the FeSe subsystem: as expected, there is a limit to the electron transfer and beyond 50% vacancy concentration no further electrons dope the FeSe. Hence, we conclude that removal of the hole pocket at Γ requires some type of surface reconstruction beyond the 1x1 1L-$TiO_2$ structure. One possibility [4] is to have 50% O vacancies and a 2x1 reconstruction of the 1L-$TiO_2$ surface which removes the hole pocket at Γ. However, the experimentally observed STO surface corresponds to a 2L-$TiO_2$ reconstruction which is also effective at removing the hole pocket with 50% O vacancies.

| Surface unit cell | % Vacancy concentration | Vacancy formation energy (eV) | |
|---|---|---|---|
|  |  | 1L $TiO_2$ | 2L $TiO_2$ |
| 1x1 | 50 | 0 | -0.22 |
| c(2x2) | 50 | -0.01 | -0.17 |
| 2x2 | 12.5 | -0.70 | -1.54 |

Table 1: Relative oxygen vacancy formation energy on the surface of $SrTiO_3$ with respect to 1x1 geometry as a function of surface unit cell and vacancy concentration.

Energetically, it is more favorable to have high oxygen vacancy concentrations in the 2L-$TiO_2$ structure compared to the 1L-$TiO_2$ termination. Table 1 shows the DFT-computed vacancy formation energy (VFE), which is the energy cost of removing an oxygen atom from the STO surface and moving it far into the vacuum. In every case, the VFE is more favorable for the 2L termination than the 1L termination. For reference, the VFE on the 2L surface is lower than that inside the STO films by 0.25 eV.

To check consistency between ARPES and DFT results for electron transfer, we have integrated the DFT densities of states to find that 0.26 electrons per Fe are needed to fill the hole pocket at Gamma of an FeSe monolayer. These electrons fill the hole pocket at Γ and also the electron pocket about M. The DFT band structure shows that the Fermi wave vectors for both pockets are very close in magnitude, so we expect ~0.13 electrons per Fe for each pocket separately. We are gratified that prior high-resolution ARPES measurements on identically prepared samples [5] estimate ~0.12 electrons per Fe from measurement of the Fermi surface about the M point.

In bulk FeSe and other iron pnictides, the bond angle X-Fe-X (X= As, Se, …) and the X-Fe height control the strength of electron-electron correlations and other properties such as Tc [52-54]. It has been shown when the As-Fe-As bond angle ranges from 106° to 114°, the Tc is greatly enhanced in various pnictide superconductors [55]. For 1UC FeSe/STO, the Se-Fe-Se bond angle in our DFT calculation is 115.1°, compared with the experimental results of 111.9° [51]. Both values are close to the range of 106° - 114°, within which the electron-electron correlations are considered to be important [55], which may explain the difference between our DFT and the experimental values.

The most straightforward consequence of the structure determined here is that the electron transfer to the FeSe is enhanced by the 2L-$TiO_2$ termination more strongly than by other surface structures of STO. The enhanced electron transfer completely suppresses the hole pocket near the Γ point and only Fermi-level crossings near the M point remain (Fig. 2(b)), which is the band structure that is generally believed to be a requirement [5, 28, 29] for superconducting 1UC FeSe/STO. Doping of the FeSe is an important aspect of the physics of FeSe/STO heterostructures. In Fe-based SCs, as well as in the cuprates, the superconducting phase competes with magnetic phases. For example, based on photoemission measurements, spin density waves (SDWs) in thin FeSe films are suppressed by heavy electron doping, promoting the superconducting phase [5]. Several mechanisms have been proposed to explain the enhanced superconductivity in 1UC FeSe/STO. An optimized doping level in FeSe may be achieved due to the stronger bonding and interfacial electron transfer. This tighter STO-FeSe coupling may also impose a more robust epitaxial constraint that could lead to a structure predicted to open strong electron-phonon coupling channels not present in the bulk [31].

To summarize, we identify the existence of double $TiO_2$ layers at the surface of STO within 1UC FeSe/STO films. The double $TiO_2$ layers play a key role in the epitaxial growth of FeSe, facilitating electron transfer between FeSe and STO and suppressing the hole pocket at the Γ point. This work provides compelling evidence that the electron transfer facilitated by the 2L-$TiO_2$ termination is critical for superconductivity in 1 UC FeSe/ $SrTiO_3$.

This research is sponsored by the DMR NSF MRSEC 1119826 and the AFOSR under Grant No. FA9550-15-1-0472. R.P., Y.P. and D.F. acknowledge support from the National Science Foundation of China, and National Basic Research Program of China (973 Program) under grant NO. 2012CB921402. The work at Argonne National Laboratory is supported by the U.S. Department of Energy. Magnetic measurements (X.H. and I.B.) at BNL were supported by the U.S. Department of Energy, Basic Energy Sciences, Materials Sciences and Engineering Division. Computational facilities are supported by NSF Grant No. CNS 08-21132 and by the facilities and staff of the Yale University Faculty of Arts and Sciences High Performance Computing Center. Additional computations are carried out via the NSF XSEDE resources through Grant No. TG-MCA08X007.

[1] F.-C. Hsu, J.-Y. Luo, K.-W. Yeh, T.-K. Chen, T.-W. Huang, P.M. Wu, Y.-C. Lee, Y.-L. Huang, Y.-Y. Chu, D.-C. Yan, M.-K. Wu, Superconductivity in the PbO-type structure alpha-FeSe, Proceedings of the National Academy of Sciences of the United States of America, 105 (2008) 14262-14264.

[2] Y. Zhang, L.X. Yang, M. Xu, Z.R. Ye, F. Chen, C. He, H.C. Xu, J. Jiang, B.P. Xie, J.J. Ying, X.F. Wang, X.H. Chen, J.P. Hu, M. Matsunami, S. Kimura, D.L. Feng, Nodeless superconducting gap in $A_xFe_2Se_2$ (A = K, Cs) revealed by angle-resolved photoemission spectroscopy, Nature Materials, 10 (2011) 273-277.

[3] Y. Lubashevsky, E. Lahoud, K. Chashka, D. Podolsky, A. Kanigel, Shallow pockets and very strong coupling superconductivity in $FeSe_xTe_{1-x}$, Nature Physics, 8 (2012) 309-312.

[4] J. Bang, Z. Li, Y.Y. Sun, A. Samanta, Y.Y. Zhang, W.H. Zhang, L.L. Wang, X. Chen, X.C. Ma, Q.K. Xue, S.B. Zhang, Atomic and electronic structures of single-layer FeSe on $SrTiO_3$(001): The role of oxygen deficiency, Physical Review B, 87 (2013) 220503.

[5] S. Tan, Y. Zhang, M. Xia, Z. Ye, F. Chen, X. Xie, R. Peng, D. Xu, Q. Fan, H. Xu, J. Jiang, T. Zhang, X. Lai, T. Xiang, J. Hu, B. Xie, D. Feng, Interface-induced superconductivity and strain-dependent spin density waves in $FeSe/SrTiO_3$ thin films, Nature Materials, 12 (2013) 634-640.

[6] I. Bozovic, C. Ahn, A new frontier for superconductivity, Nature Physics, 10 (2014) 892-895.

[7] S. Mandal, R.E. Cohen, K. Haule, Strong pressure-dependent electron-phonon coupling in FeSe, Physical Review B, 89 (2014) 220502.

[8] R. Peng, X.P. Shen, X. Xie, H.C. Xu, S.Y. Tan, M. Xia, T. Zhang, H.Y. Cao, X.G. Gong, J.P. Hu, B.P. Xie, D.L. Feng, Measurement of an Enhanced Superconducting Phase and a Pronounced Anisotropy of the Energy Gap of a Strained FeSe Single Layer in $FeSe/Nb:SrTiO_3/KTaO_3$ Heterostructures Using Photoemission Spectroscopy, Physical Review Letters, 112 (2014) 107001.

[9] R. Peng, H.C. Xu, S.Y. Tan, H.Y. Cao, M. Xia, X.P. Shen, Z.C. Huang, C.H.P. Wen, Q. Song, T. Zhang, B.P. Xie, X.G. Gong, D.L. Feng, Tuning the band structure and superconductivity in single-layer FeSe by interface engineering, Nature Communications, 5 (2014).

[10] Y. Sun, W. Zhang, Y. Xing, F. Li, Y. Zhao, Z. Xia, L. Wang, X. Ma, Q.-K. Xue, J. Wang, High temperature superconducting FeSe films on $SrTiO_3$ substrates, Scientific Reports, 4 (2014) 6040.

[11] Q. Fan, W.H. Zhang, X. Liu, Y.J. Yan, M.Q. Ren, R. Peng, H.C. Xu, B.P. Xie, J.P. Hu, T. Zhang, D.L. Feng, Plain s-wave superconductivity in single-layer FeSe on $SrTiO_3$ probed by scanning tunnelling microscopy, Nature Physics, 11 (2015) 946-952.

[12] J.-F. Ge, Z.-L. Liu, C. Liu, C.-L. Gao, D. Qian, Q.-K. Xue, Y. Liu, J.-F. Jia, Superconductivity above 100 K in single-layer FeSe films on doped $SrTiO_3$, Nature Materials, 14 (2015) 285-289.

[13] D. Huang, C.-L. Song, T.A. Webb, S. Fang, C.-Z. Chang, J.S. Moodera, E. Kaxiras, J.E. Hoffman, Revealing the Empty-State Electronic Structure of Single-Unit-Cell $FeSe/SrTiO_3$, Physical Review Letters, 115 (2015) 017002.

[14] M.K. Wu, P.M. Wu, Y.C. Wen, M.J. Wang, P.H. Lin, W.C. Lee, T.K. Chen, C.C. Chang, An overview of the Fe-chalcogenide superconductors, Journal of Physics D-Applied Physics, 48 (2015).

[15] S. Lee, C. Tarantini, P. Gao, J. Jiang, J.D. Weiss, F. Kametani, C.M. Folkman, Y. Zhang, X.Q. Pan, E.E. Hellstrom, D.C. Larbalestier, C.B. Eom, Artificially engineered superlattices of pnictide superconductors, Nature Materials, 12 (2013) 392-396.

[16] Y. Mizukami, M. Konczykowski, Y. Kawamoto, S. Kurata, S. Kasahara, K. Hashimoto, V. Mishra, A. Kreisel, Y. Wang, P.J. Hirschfeld, Y. Matsuda, T. Shibauchi, Disorder-induced topological change of the superconducting gap structure in iron pnictides, Nature Communications, 5 (2014).

[17] X.H. Chen, T. Wu, G. Wu, R.H. Liu, H. Chen, D.F. Fang, Superconductivity at 43 K in $SmFeAsO_{(1-x)}F_x$, Nature, 453 (2008) 761-762.

[18] Y. Kamihara, T. Watanabe, M. Hirano, H. Hosono, Iron-based layered superconductor $LaO_{1-x}F_xFeAs$ (x=0.05-0.12) with $T_c$=26 K, Journal of the American Chemical Society, 130 (2008) 3296.

[19] Y. Mizuguchi, F. Tomioka, S. Tsuda, T. Yamaguchi, Y. Takano, Superconductivity at 27 K in tetragonal FeSe under high pressure, Applied Physics Letters, 93 (2008) 152505.
[20] T. Imai, K. Ahilan, F.L. Ning, T.M. McQueen, R.J. Cava, Why Does Undoped FeSe Become a High-$T_c$ Superconductor under Pressure?, Physical Review Letters, 102 (2009) 177005.
[21] T.M. McQueen, A.J. Williams, P.W. Stephens, J. Tao, Y. Zhu, V. Ksenofontov, F. Casper, C. Felser, R.J. Cava, Tetragonal-to-Orthorhombic Structural Phase Transition at 90 K in the Superconductor $Fe_{1.01}Se$, Physical Review Letters, 103 (2009) 057002.
[22] S. Medvedev, T.M. McQueen, I.A. Troyan, T. Palasyuk, M.I. Eremets, R.J. Cava, S. Naghavi, F. Casper, V. Ksenofontov, G. Wortmann, C. Felser, Electronic and magnetic phase diagram of beta-$Fe_{1.01}Se$ with superconductivity at 36.7 K under pressure, Nature Materials, 8 (2009) 630-633.
[23] J. Guo, S. Jin, G. Wang, S. Wang, K. Zhu, T. Zhou, M. He, X. Chen, Superconductivity in the iron selenide $K_xFe_2Se_2$ (0 <= x <= 1.0), Physical Review B, 82 (2010) 180520.
[24] D. Liu, W. Zhang, D. Mou, J. He, Y.-B. Ou, Q.-Y. Wang, Z. Li, L. Wang, L. Zhao, S. He, Y. Peng, X. Liu, C. Chen, L. Yu, G. Liu, X. Dong, J. Zhang, C. Chen, Z. Xu, J. Hu, X. Chen, X. Ma, Q. Xue, X.J. Zhou, Electronic origin of high-temperature superconductivity in single-layer FeSe superconductor, Nature Communications, 3 (2012) 931.
[25] Q.-Y. Wang, Z. Li, W.-H. Zhang, Z.-C. Zhang, J.-S. Zhang, W. Li, H. Ding, Y.-B. Ou, P. Deng, K. Chang, J. Wen, C.-L. Song, K. He, J.-F. Jia, S.-H. Ji, Y.-Y. Wang, L.-L. Wang, X. Chen, X.-C. Ma, Q.-K. Xue, Interface-Induced High-Temperature Superconductivity in Single Unit-Cell FeSe Films on $SrTiO_3$, Chinese Physics Letters, 29 (2012) 037402.
[26] Y.-Y. Xiang, F. Wang, D. Wang, Q.-H. Wang, D.-H. Lee, High-temperature superconductivity at the FeSe/$SrTiO_3$ interface, Physical Review B, 86 (2012) 134508.
[27] J. Bang, Z. Li, Y.Y. Sun, A. Samanta, Y.Y. Zhang, W. Zhang, L. Wang, X. Chen, X. Ma, Q.K. Xue, S.B. Zhang, Atomic and electronic structures of single-layer FeSe on $SrTiO_3$(001): The role of oxygen deficiency, Physical Review B, 87 (2013) 220503.
[28] S. He, J. He, W. Zhang, L. Zhao, D. Liu, X. Liu, D. Mou, Y.-B. Ou, Q.-Y. Wang, Z. Li, L. Wang, Y. Peng, Y. Liu, C. Chen, L. Yu, G. Liu, X. Dong, J. Zhang, C. Chen, Z. Xu, X. Chen, X. Ma, Q. Xue, X.J. Zhou, Phase diagram and electronic indication of high-temperature superconductivity at 65 K in single-layer FeSe films, Nature Materials, 12 (2013) 605-610.
[29] J.J. Lee, F.T. Schmitt, R.G. Moore, S. Johnston, Y.T. Cui, W. Li, M. Yi, Z.K. Liu, M. Hashimoto, Y. Zhang, D.H. Lu, T.P. Devereaux, D.H. Lee, Z.X. Shen, Interfacial mode coupling as the origin of the enhancement of $T_c$ in FeSe films on $SrTiO_3$, Nature, 515 (2014) 245-U207.
[30] W.-H. Zhang, Y. Sun, J.-S. Zhang, F.-S. Li, M.-H. Guo, Y.-F. Zhao, H.-M. Zhang, J.-P. Peng, Y. Xing, H.-C. Wang, T. Fujita, A. Hirata, Z. Li, H. Ding, C.-J. Tang, M. Wang, Q.-Y. Wang, K. He, S.-H. Ji, X. Chen, J.-F. Wang, Z.-C. Xia, L. Li, Y.-Y. Wang, J. Wang, L.-L. Wang, M.-W. Chen, Q.-K. Xue, X.-C. Ma, Direct Observation of High-Temperature Superconductivity in One-Unit-Cell FeSe Films, Chinese Physics Letters, 31 (2014) 017401.
[31] S. Coh, M.L. Cohen, S.G. Louie, Large electron-phonon interactions from FeSe phonons in a monolayer, New Journal of Physics, 17 (2015) 73027-73027.
[32] Y. Miyata, K. Nakayama, K. Sugawara, T. Sato, T. Takahashi, High-temperature superconductivity in potassium-coated multilayer FeSe thin films, Nature Materials, 14 (2015) 775.
[33] F.S. Li, H. Ding, C.J. Tang, J.P. Peng, Q.H. Zhang, W.H. Zhang, G.Y. Zhou, D. Zhang, C.L. Song, K. He, S.H. Ji, X. Chen, L. Gu, L.L. Wang, X.C. Ma, Q.K. Xue, Interface-enhanced high-temperature superconductivity in single-unit-cell $FeTe_{1-x}Se_x$ films on $SrTiO_3$, Physical Review B, 91 (2015) 220503.
[34] X. Liu, D. Liu, W. Zhang, J. He, L. Zhao, S. He, D. Mou, F. Li, C. Tang, Z. Li, L. Wang, Y. Peng, Y. Liu, C. Chen, L. Yu, G. Liu, X. Dong, J. Zhang, C. Chen, Z. Xu, X. Chen, X. Ma, Q. Xue, X.J. Zhou,

Dichotomy of the electronic structure and superconductivity between single-layer and double-layer FeSe/SrTiO$_3$ films, Nature Communications, 5 (2014).
[35] J.H. Tapp, Z.J. Tang, B. Lv, K. Sasmal, B. Lorenz, P.C.W. Chu, A.M. Guloy, LiFeAs: An intrinsic FeAs-based superconductor with T$_c$=18 K, Physical Review B, 78 (2008) 060505.
[36] X.C. Wang, Q.Q. Liu, Y.X. Lv, W.B. Gao, L.X. Yang, R.C. Yu, F.Y. Li, C.Q. Jin, The superconductivity at 18 K in LiFeAs system, Solid State Communications, 148 (2008) 538-540.
[37] K. Ishida, Y. Nakai, H. Hosono, To What Extent Iron-Pnictide New Superconductors Have Been Clarified: A Progress Report, Journal of the Physical Society of Japan, 78 (2009) 062001.
[38] E. Dagotto, Correlated Electrons In High-Temperature Superconductors, Reviews of Modern Physics, 66 (1994) 763-840.
[39] P.A. Lee, N. Nagaosa, X.G. Wen, Doping a Mott insulator: Physics of high-temperature superconductivity, Reviews of Modern Physics, 78 (2006) 17-85.
[40] A.P. Drozdov, M.I. Eremets, I.A. Troyan, V. Ksenofontov, S.I. Shylin, Conventional superconductivity at 203 kelvin at high pressures in the sulfur hydride system, Nature, 525 (2015) 73-+.
[41] C.L. Song, Y.L. Wang, P. Cheng, Y.P. Jiang, W. Li, T. Zhang, Z. Li, K. He, L.L. Wang, J.F. Jia, H.H. Hung, C.J. Wu, X.C. Ma, X. Chen, Q.K. Xue, Direct Observation of Nodes and Twofold Symmetry in FeSe Superconductor, Science, 332 (2011) 1410-1413.
[42] C.L. Song, Y.L. Wang, Y.P. Jiang, Z. Li, L.L. Wang, K. He, X. Chen, X.C. Ma, Q.K. Xue, Molecular-beam epitaxy and robust superconductivity of stoichiometric FeSe crystalline films on bilayer graphene, Physical Review B, 84 (2011) 020503.
[43] H. Ding, Y.F. Lv, K. Zhao, W.L. Wang, L.L. Wang, C.L. Song, X. Chen, X.C. Ma, Q.K. Xue, High-temperature superconductivity in single-unit-cell FeSe films on anatase TiO$_2$(001), arXiv:1603.00999 (2016).
[44] M. Kareev, S. Prosandeev, J. Liu, C. Gan, A. Kareev, J.W. Freeland, M. Xiao, J. Chakhalian, Atomic control and characterization of surface defect states of TiO$_2$ terminated SrTiO$_3$ single crystals, Applied Physics Letters, 93 (2008).
[45] See Supplemental Material at http://link.aps.org/supplemental/... for a description of further details related to sample preparation methods, density dunctional theory calculations, RHEED of FeSe grown on 1x1 1L-TiO$_2$ terminated SrTiO$_3$, crystal truncation rods of SrTiO$_3$, line cuts of RHEED measurements, electron transfer and density of states at FeSe/SrTiO$_3$ interfaces.
[46] N. Erdman, L.D. Marks, SrTiO$_3$(001) surface structures under oxidizing conditions, Surface Science, 526 (2003) 107-114.
[47] O. Warschkow, M. Asta, N. Erdman, K.R. Poeppelmeier, D.E. Ellis, L.D. Marks, TiO$_2$-rich reconstructions of SrTiO$_3$(001): a theoretical study of structural patterns, Surface Science, 573 (2004) 446-456.
[48] D.M. Kienzle, A.E. Becerra-Toledo, L.D. Marks, Vacant-Site Octahedral Tilings on SrTiO$_3$ (001), the (root 13 x root 13)R33.7 degrees Surface, and Related Structures, Physical Review Letters, 106 (2011) 176102.
[49] H.Y. Cao, S.Y. Tan, H.J. Xiang, D.L. Feng, X.G. Gong, Interfacial effects on the spin density wave in FeSe/SrTiO$_3$ thin films, Physical Review B, 89 (2014) 014501.
[50] K.V. Shanavas, D.J. Singh, Doping SrTiO$_3$ supported FeSe by excess atoms and oxygen vacancies, Physical Review B, 92 (2015) 035144.
[51] L. Fangsen, Z. Qinghua, T. Chenjia, L. Chong, S. Jinan, N. CaiNa, Z. Guanyu, L. Zheng, Z. Wenhao, S. Can-Li, H. Ke, J. Shuaihua, Z. Shengbai, G. Lin, W. Lili, M. Xu-Cun, X. Qi-Kun, Atomically resolved FeSe/SrTiO$_3$(001) interface structure by scanning transmission electron microscopy, 2D Materials, 3 (2016) 024002.

[52] H. Okabe, N. Takeshita, K. Horigane, T. Muranaka, J. Akimitsu, Pressure-induced high-$T_c$ superconducting phase in FeSe: Correlation between anion height and $T_c$ Physical Review B, 81 (2010) 205119.
[53] F. Essenberger, P. Buczek, A. Ernst, L. Sandratskii, E.K.U. Gross, Paramagnons in FeSe close to a magnetic quantum phase transition: *Ab initio* study, Physical Review B, 86 (2012) 060412.
[54] C. Zhang, L.W. Harriger, Z. Yin, W. Lv, M. Wang, G. Tan, Y. Song, D.L. Abernathy, W. Tian, T. Egami, K. Haule, G. Kotliar, P. Dai, Effect of Pnictogen Height on Spin Waves in Iron Pnictides, Physical Review Letters, 112 (2014) 217202.
[55] C.H. Lee, A. Iyo, H. Eisaki, H. Kito, M.T. Fernandez-Diaz, T. Ito, K. Kihou, H. Matsuhata, M. Braden, K. Yamada, Effect of structural parameters on superconductivity in fluorine-free LnFeAsO$_{1-y}$ (Ln = La, Nd), Journal of the Physical Society of Japan, 77 (2008).